\documentclass[11pt,twoside]{article}
\usepackage{asp2010}
\usepackage{url}
\resetcounters

\begin{document}

\title{Colour-magnitude diagrams, probabilistic synthesis models and 
the upper mass limit of the initial mass function}
\author{Miguel Cervi{\~n}o$^1$, Enrique P{\'e}rez$^1$, Nestor S{\'a}nchez$^1$, Carlos Rom{\'a}n-Z{\'u}{\~n}iga$^{1,2}$ and David Valls-Gabaud$^3$
\affil{$^1$Instituto de Astrof{\'\i}sica de Andaluc{\'\i}a (IAA-CSIC), Glorieta de la Astronom{\'\i}a, 18008 Granada, Spain}
\affil{$^2$Centro Astron\'omico Hispano Alem\'an, 18006 Granada, Spain}
\affil{$^3$GEPI, CNRS UMR 8111, Observatoire de Paris, 5 Place Jules Janssen, 92195 Meudon, France}}

\begin{abstract}
We present the underlying relations between colour-magnitude diagrams (CMDs) and synthesis models through the use of stellar luminosity distribution functions. CMDs studies make a direct use of the stellar luminosity distribution function while, in general, synthesis models only use its mean value, even though high-order moments can also be obtained. We show that the mean, high-order moments and integrated luminosity distribution functions of stellar ensembles are related to the stellar luminosity distribution function, within the formalism of probabilistic synthesis models. More details have been yet presented
in \cite{CLCL06} and references therein.
As a direct application of this formalism, we discuss  two key issues. First, inferences on the upper mass limit
of the initial mass function as a function of the total mass of clusters. Second, we apply
extreme value theory  to show that that the cluster mass obtained from normalising  the IMF between $m_\mathrm{max}$  and $m_\mathrm{up}$ does not provide the cluster mass in the case where
 only one star in this mass range is present, as assumed in the IGIMF theory. It provides instead 
the cluster mass with a 60\% probability to have a star with mass larger than $m_\mathrm{max}$, and
we argue that in light of this result the basic formulation ofthe IGIMF theory must be revised.
\end{abstract}

\section{From stars to stellar ensembles and the mass-luminosity relation}

Our basic knowledge of the Universe stems from the light received from observed sources. In a first-order approximation (neglecting interactions with the interstellar medium and non-stellar components), we can consider two types of sources: individual stars and stellar ensembles. Since in this case the emission of an stellar ensemble is just the sum of its individual components, we can refer to this emission as {\it integrated} light, coming from an unresolved
system such as a distant cluster, or a pixel/slit/IFU in an image of a galaxy. The problem of inferring physical properties from the observed data can thus be reduced to the analysis of the observed light of individual stars in terms of theoretical stellar models, or, in the case of integrated light, to decompose the integrated light into its (stellar) components.

Obviously, the interpretation and physical inferences that can be obtained from the integrated light depends on our physical knowledge of the individual sources that would be present in the ensemble. It implies to have theoretical models which cover the emission of all possible stars in the ensemble or, at least, to be able to model the emission of the most luminous ones, which are also the more massive in a first order approximation. The reason being that they dominate the integrated emission of stellar ensembles. However, the evolution and spectra of 
massive/luminous stars are far from being  solved problems in astronomy, and constitute a very active research area. The impact of mass loss, rotation, magnetic fields, binary interactions, etc, can change the physical inferences obtained from the observed data  \cite[see the 
contribution by][in these proceedings]{Mass10}.

The fact that integrated light  is  dominated  by the most luminous stars could be an advantage for some studies dealing with age inferences.
It is, however, a problem if we are interested in inferring the total mass as the statistics of stars present in the field \cite[e.g.][]{Sal55} show that low-luminosity, low-mass stars are the most numerous ones and dominate the total mass budget of any ensemble, while high-mass stars are the less numerous, with a small contribution to the total mass, but dominate the integrated light.

The problem we are addressing here, then, is the inference about the total mass of the ensemble given its total, integrated luminosity. Homology relations in zero-age main sequence stars allow to relate the current stellar mass $m(t)$ and its bolometric luminosity \cite[which is extrapolated to the luminosity in any band, wavelength or time $\ell_{\lambda}(t)$ in synthesis codes, see][ and references therein for a discussion]{CL05} by a power law as

\begin{equation}
\ell_{\lambda}(t) \propto m(t)^{\gamma} \; ,
\label{eq:ml_stars}
\end{equation}

\noindent where  $\gamma$ has typical value of around $3$ for main-sequence stars. 

In the case of {\it unresolved} stellar ensembles we are seeking a mass-luminosity relation with an even simpler functional form:

\begin{equation}
{\cal{L}_{\lambda}}(t)/{\cal{M}}(t) \propto \mathrm{constant} \; , 
\label{eq:ml_ensemble}
\end{equation}

\noindent where ${\cal{L}_{\lambda}}(t)$ the integrated luminosity in a given band/wavelength of an stellar ensemble whose total mass in stars is ${\cal{M}}(t)$. We stress the difference in the  functional dependence of the luminosity in Eq. \ref{eq:ml_stars} and Eq. \ref{eq:ml_ensemble}.  To obtain these simple relations was actually the problem that Beatrice
Tinsley aimed to solve \cite[e.g.][]{Tin80}, and it leads to the development of stellar population synthesis models.

\section{Stellar evolution, CMDs and {\it probabilistic} synthesis models}

Stellar evolution theory (in particular evolutionary tracks) predicts the evolution of stars with a given initial mass $m$ as a function of time $t$ measured from the so-called zero-age main sequence. We note this as,
for example, the luminosity of a star as a function of time, {\it given}
its mass: $\ell_\mathrm{track}(t|m)$. During the main sequence phase, the mass-luminosity relation is more or less well described by Eq. \ref{eq:ml_stars}, although such a relation does not apply for evolved phases. Provided a large grid of  tracks and an interpolation algorithm we can transform the luminosity of a set of stars with given initial masses as a function of time, $\ell_\mathrm{track}(t|m)$,  into the evolution of stars at a {\it given} time as a function of their initial masses, $\ell_\mathrm{iso}(m|t)$, that is, isochrones    \cite[such transformation is not always a trivial task, we refer to, e.g.,][ for details]{CL05}.

At first glance, Colour-Magnitude Diagrams (CMDs) are just a 
projection of such isochrones in a particular plane given by the choice
of filters. Until recently, the comparison of the {\it shape} of the observed sequences in these diagrams with a set of isochrones provided highly valuable information about physical properties of the ensemble of stars, such as ages, distances and metallicities. The comparison of shapes, however, results in a number of degeneracies, such as the age-metallicity
degeneracy in the sense that isochrones of different ages selected at
different metallicities have the same {\it shape}. One has to keep in
mind that, besides the observational errors, the location of stars in
a CMD is a sequence of {\it mass}. When a CMD is not just considered as the 2-D locus of stars of a given age and metallicity, but as a density structure which takes into account {\it how many} stars are in each point of the diagram, the degeneracy can be  lifted. The reason is that the density of stars {\it along} an isochrone is a very sensitive probe 
of both age and metallicity. 
So, taken into account the density,
not only can the basic parameters of these single stellar populations
be inferred \cite[see][for in a classical Bayesian way]{Hernandez:2008qy}, but also
their star formation and chemical evolution histories \cite[see, for a review, ][]{dvg2011}.

A similar formalism for interpreting  CMDs taking into account the density is the representation of the stellar luminosity distribution function (the distribution of stars that share a common, given luminosity). Taking the distribution that provides how many stars are born with a given mass $m$ at a given age $t$, that is,
the stellar birth-rate  $b(m,t|{\cal M})$, any stellar luminosity distribution function or CMD can be produced. The aim of CMD analyses 
is just the inverse problem, that is, to infer $b(m,t)$ from observations.

It is traditionally assumed \cite[see, e.g.,][]{Tin80} that the stellar birth rate can be decomposed into two  {\it separable} functions: the 
stellar initial mass function (IMF) $\phi(m)$, and the total amount of gas transformed into stars at a given time, i.e., the star formation rate
history $\psi({\cal M},t) = d{\cal M}(t)/dt$:
\begin{equation}
b(m,t| {\cal M}) \, \mathrm{d}m \, \mathrm{d}t =  \phi(m) \, \mathrm{d}m \times 
\psi(t|{\cal M})\,\mathrm{d}t  \; .
\end{equation}
This decomposition of the stellar birth rate is the basis of most of the actual research in the studies of stellar ensembles: the use of isochrones make sense only under the assumption that the stellar masses produced in an star formation event do not depend on the amount of gas transformed into stars in the event itself nor in previous events. Formally, this decomposition with small modifications remains valid if an initial cluster mass function, ICMF $\varphi({\cal{M}}_\mathrm{clus})$ is included \citep{WKB10}:
\begin{equation}
b(m,{\cal{M}}_\mathrm{clus},t| {\cal M}) \, \mathrm{d}m \, \mathrm{d}{\cal{M}}_\mathrm{clus}\, \mathrm{d}t =  \phi(m) \, \mathrm{d}m \times \varphi({\cal{M}}_\mathrm{clus}) \, \mathrm{d}{\cal{M}}_\mathrm{clus} \times \psi(t|{\cal M}_\mathrm{clus})\,\mathrm{d}t \; ,
\end{equation}
\noindent as far as an integrated galactic initial mass function, IGIMF,  which includes the IMF and ICMF functions, can be separated from the star formation  history, and the star formation history refers to the amount of gas transformed into clusters, ${\cal M}_\mathrm{clus}$,  instead of the amount of gas transformed into stars, ${\cal M}$.

In the following we just consider the IMF instead the IGIMF to obtain the stellar luminosity function.
Given the IMF and the isochrone at a given time we can obtain the distribution function that gives us the probability of obtain a star with a given luminosity. The implicit assumption is that the IMF is a probability distribution function which provides the probability to have a star with a given mass, but {\sl not} an exact number of stars. The stellar luminosity function $\varphi_\ell \, \mathrm{d}\ell$ is, trivially, \citep{CLCL06} 

\begin{equation}
\varphi_\ell = \phi(m) \times \left(\frac{\mathrm{d}\ell(m)}{\mathrm{d}m}\right)^{-1} \, ,
\end{equation}

The different inflection points in the mass-luminosity relation will yield
features in the luminosity function which do not appear in the mass function.
The {\it mean} of the distribution (i.e. the mean luminosity of clusters which contain just one star, $\left< {\cal{L}}_{1*} \right>$) is 

\begin{equation}
\left< {\cal{L}}_{1*} \right> = \int_{m_\mathrm{low}}^{m_\mathrm{up}} \ell(m)\, \phi(m)  \left(\frac{\mathrm{d}\ell(m)}{\mathrm{d}m}\right)^{-1}  \, \frac{\mathrm{d}\ell(m)}{\mathrm{d}m} \, \mathrm{d}m =  \int_{m_\mathrm{low}}^{m_\mathrm{up}} \ell(m)\, \phi(m) \, \mathrm{d}m \, ,
\label{eq:lmean}
\end{equation}

\noindent which is the equation solved by synthesis codes. 

Equation \ref{eq:lmean} shows two important issues that must be considered with caution. The first one is related with the derivative of $\ell$ with the initial mass $m$. Such derivative is not always defined and isochrones have discontinuities in post-main sequence evolutionary phases, the so called {\it fast evolutionary phases} where Eq. \ref{eq:ml_stars} does not apply. 
Actually such phases are better described in terms of the lifetime of the phase for a given mass $m$ (typically the turn-off mass) rather than with initial masses. This issue leads to different algorithms to compute synthesis models, either  using {\it fuel consumption theorem} or else isochrone synthesis \cite[see][ for more details]{Buzz89,MG01}. The important point here is that since it  is the stellar luminosity distribution function the one that must be well sampled, size-of-sample effects may be still present in the sampling of fast evolutionary phases, {\sl even with a well-sampled IMF}.

The second cautionary note refers to the IMF mass limits themselves. The formulation and the modelisation of stellar ensembles is valid as far as $m_\mathrm{up}$ (and formally $m_\mathrm{low}$) have  well-defined values. The mean value depends on both parameters and the functional form of the stellar luminosity distribution function.

When dealing with an ensemble of $N$ stars the corresponding distribution which describes the possible integrated luminosities of the ensemble is $N$ times the self-convolution of the stellar luminosity function (under random sampling of the stellar luminosity distribution function assumption). Then the mean total luminosity is \cite[see][ for details]{CLCL06}:

\begin{equation}
\left< {\cal{L}}_{N*} \right> = N \times  \left< {\cal{L}}_{1*} \right> = N \times  \left< \ell \right> \, .
\end{equation}

\noindent We stress again that, as far as ({\sl i}) the stellar luminosity distribution function must be sampled and {\sl (ii)} its high luminosity tail is related to the lifetime of particular evolutionary phases, a random sampling hypothesis (for the stellar luminosity function) is still valid even under a sorted sampled IMF hypothesis.

This formalism can equally be applied to the total mass of the ensemble: the total mass distribution of an ensemble of $N$ stars is described by  $N$ self-convolutions of the IMF itself \citep{SM08} and the mean value of the total mass of the ensemble is $N$ times the mean mass of the IMF:

\begin{equation}
\left< {\cal{M}}_{N*} \right> = N \times  \left< {\cal{M}}_{1*} \right> = N \times \left< m \right> \, .
\end{equation}

Remarquably, this implies that

\begin{equation}
\frac{\left< {\cal{L}}_{N*} \right>}{\left< {\cal{M}}_{N*} \right>} =\frac{N \times \left< {\cal{L}}_{1*} \right>}{N \times \left< {\cal{M}}_{1*} \right>} = \frac{\left< {\cal{L}}_{1*} \right>}{ \left< {\cal{M}}_{1*} \right>} = \mathrm{constant}
\end{equation}
\noindent and hence validates the approach given by Eq. \ref{eq:ml_ensemble}. Note that this relation provides a valid (averaged) total mass as far as the observed luminosity is a good proxy of the mean luminosity of stellar ensembles with $N$ stars, and the age is known.

\section{Size-of-sample issues}

We have shown that synthesis models results are  related to CMD analyses 
through a collapse of the underlying distribution that defines the CMD in its moments. Obviously, the mean value of the distribution is not the only possible outcome of synthesis codes, and high-order moments or even the distributions themselves can be obtained \citep{CLCL06}. Working with distribution moments has the advantage of providing a first order approximation of the underling distribution. From the Central Limit theorem, we know that an infinite self-convolution of a distribution with finite limits tends to the  Gaussian form. High-order moments allows us to define how many stars (and average mass) are needed to obtain Gaussian or quasi-Gaussian integrated luminosity distributions \cite[see][ for details]{CLCL06}. This is the basic requirement for $\chi^2$ fitting algorithms\footnote{Although $\chi^2$ seems the simplest tool to make inferences from observed data (the inverse problem) we emphasize that there are many different methodologies (as well as interpretations) on how to deal with inverse problems. For a simple introduction see \cite{Tarantola} as well  as other articles at \url{http://www.ipgp.fr/~tarantola}.}.
In particular the second moment of the stellar luminosity distribution function (the variance) also allows one to evaluate the relative error in the use of an observation as a proxy of the mean of the underling distribution. However the situation becomes more complicated when a {\it small} number of stars  defines the integrated luminosity\footnote{Note that {\it small} is not just an intrinsic property of the studied system, but also an observational effect: an IFU observation intrinsically 
contains a smaller number of stars than a large aperture observation.}.

In general three cases can be defined: ({\sl a}) Systems/observations which contain a large number of stars yielding gaussian distributions of the theoretical integrated luminosity. In this case the observations can be used as proxy of mean values safely. These situations typically correspond to galaxies and systems  with masses larger than $10^8$ M$_\odot$ when all the light from the system is covered by the IFU/slit. ({\sl b}) Systems/observations with a large number of stars to ensure gaussian-like distributions, but not large enough to provide a negligible variance. In these cases both the mean value and  the variance of the underling distribution for each band/wavelength {\sl must} be considered.
This situation corresponds to systems typically more massive than $10^5$ M$_\odot$ although the exact value depends on the considered band/wavelength. ({\sl c}) {\it Extreme size-of-sample effects}: these are systems/observations with a number of stars so small that the distributions of the integrated luminosities are not Gaussians. In these cases, the analysis of observed data in terms of the mean and high-order moments of the distribution of the integrated luminosity is {\sl not} suitable since they do not provide a proper information on the distribution (mean and mode differs, variance can be not translated in confidence intervals etc.). In this situation, it is more accurate to work directly with integrated luminosity distribution functions which  can be obtained by self-convolution of the stellar luminosity distribution function (which has several problems at computational level except for a low number of stars) or by the analysis of Monte Carlo simulations (which are more economic at the computational level but which lead to some subtle problems, such as increasing the number of simulations when the number of stars in the simulated clusters decreases to provide a correct sampling of the underling distributions).

\section{Low Luminosity Limit (LLL), ${\cal{M}}_\mathrm{clus}$ and $m_\mathrm{up}$}

Unfortunately we have no estimation of the total mass of the system before an analysis is performed, and simple recipes are useful to know in which of the previous r\'egimes the observational data lie. A simple method, called the Low Luminosity Limit (LLL), was proposed by \cite{CL04}. The method just compares the luminosity of the observed ensemble with the luminosity of the most luminous star assumed in the model, that is, the extreme value of the stellar luminosity distribution function (the LLL). It provides the lowest luminosity that an stellar ensemble should have without been
possibly confused with an single star.

An additional simple test  is just the comparison of the observed light from a stellar ensemble with the locus of the mean values obtained by synthesis models and the locus of individual stars in color-color diagrams \cite[see][as an example]{BdGC08}: clusters affected by strong sampling effects (where the mean values are not representative) cover the intermediate area between stars and the mean values obtained by synthesis models. 

A similar situation is present while making inferences of total masses ${\cal{M}}_\mathrm{clus}$ for studies of the maximum mass $m_\mathrm{max}$ that a cluster with total mass ${\cal{M}}_\mathrm{clus}$ would contain with $m_\mathrm{max} - {\cal{M}}_\mathrm{clus}$ diagrams \citep[][and references therein]{SM08,WKB10}. The observed maximum stellar mass can be obtained from observations. However, ${\cal{M}}_\mathrm{clus}$ must be inferred under a situation where we know for sure that the IMF is not well sampled. In these cases, one has to use all observational constrains (such as the number of stars with different masses), run a large enough set of Monte Carlo simulations which fulfills the observational constrains, and obtain the distribution of possible total masses. Note that the usual method of correcting the total mass in the unobserved mass range with the mass obtained from a truncated IMF from $m_\mathrm{low}$ to the minimum observed mass $m_\mathrm{min}^\mathrm{obs}$ implicitly assumes that the IMF is fully sampled in the low mass interval (which is equivalent to say that the IMF is sorted sampled up to $m_\mathrm{min}^\mathrm{obs}$).

Finally, we address the problem of the inference of $m_\mathrm{up}$  from observations. The probability of an extreme value, say $m_\mathrm{up}$, is precisely the subject of  {\it Extreme Value Theory} (EVT). This branch of statistics deals with extreme deviations (maxima or minima) 
from the median and it has important consequences in every day life (such as economic crashes) and natural catastrophes produced by deviations in annual flood flows, precipitation maxima or earthquakes (also with human-life and economic implications). The classical reference in the subject is \cite{Gum58}, although we also suggest reading \cite{Sor04} (especially pp. 18-23), where the case
 $m_\mathrm{up} = \infty$ is discussed in detail, and where results
 obtained in astronomy (e.g. \cite{vAlb69,OC05,P-AK08}) can be  also be
 found. A detailed discussion is presented in a forthcoming paper.

Within this EVT framework, we analyse the basic assumption made in the IGIMF theory proposed by \cite{WKB10} and previous works. Let us consider a sample of $N$ stars. An event with probability $p$ occurs typically $N \times p$ times, hence if we {\it expect} just 1 star in the mass range $m_\mathrm{max}-m_\mathrm{up}$  in the cluster we have:

\begin{equation}
p=\int_{m_\mathrm{max}}^{m_\mathrm{up}} \phi(m) \, \mathrm{d}m; ~~~~
 Np = 1; ~~~~~ p = \frac{1}{N}
\end{equation}

\noindent and we can obtain the total number of stars and make an estimation of the total mass of such a cluster, $ \left< {\cal{M}}_\mathrm{clus}\right> = N \times \left< m \right>$, \cite[provided that we are in a r\'egime where  $\left< m \right>$ can be safely used! but see][]{SM08}.

We also know  from EVT the PDF of the {\sl extreme} values for $N$ stars, so, we can ask what $1/N$ really means. Following \cite{Sor04} the estimation of  $m_\mathrm{max}$ from $1/N$ is in fact 
the stellar mass that is not exceeded with a probability $p = 1/6 = 0.37$ for a cluster with $N$ stars. It means that $(1-p)$=63\% of clusters with $N$ stars (and hence $\left< {\cal{M}}_\mathrm{clus}\right> = N \times \left< m \right>$) have a star with mass {\it larger} than $m_\mathrm{max}$. In other words, the estimation of the relation between $m_\mathrm{max}$ and ${\cal{M}}_\mathrm{clus}$ as assumed in the IGIMF theory may not be  correct in 63\% of real cases. 

\section{Conclusions}

In this contribution we have shown that the physical inferences obtained from CMDs studies are related with the results of evolutionary synthesis codes by the stellar luminosity distribution function. Since CMDs basically provide information about the distribution itself and the synthesis models typically just provide the mean of such distribution, a direct conclusion is that a CMD provides {\it always} a more complete information about  star formation than the use of mean values obtained by synthesis models.

The mean of the stellar luminosity distribution function is defined by its functional form as well as its limits (the upper one is non-trivially related to $m_\mathrm{up}$). This mean as well as high-order moments are directly proportional to the mean integrated luminosity of ensembles of $N$ stars. Such proportionality provides mass-luminosity relations for systems/observations with a large number of stars (such that the average total masses are larger than $10^8$~M$_\odot$) where the observed luminosity can be safely used as a proxy of the theoretical mean integrated luminosity. However, the use of one observation as a proxy of the mean of the theoretical distribution of integrated luminosities must be used with caution for less massive systems:  it must be used together with the variance for ensembles with mean mass values in the range $10^5$ to $10^8$~M$_\odot$,  and simply {\sl not used} for ensembles with mean total masses smaller than 10$^5$~M$_\odot$. For these small (by number) 
systems ones has to use the distribution of integrated luminosities 
rather the parametric descriptions of the distribution.

The reason of these limitations is the incomplete sampling (or size-of-sample effects) of the stellar luminosity distribution function in the ensemble. Since the stellar luminosity distribution function contains a (low-luminosity) component related to the IMF, and a (high luminosity) component related to the short lifetime of fast evolutionary phases, an intrinsic scatter due to random sampling is always present even in cases of a perfect or sorted sampling of the IMF. More details about these 
limitations can be found in \cite{CL04,CL05,CLCL06} and references therein.

One of the difficulties related with size-of-sample effects is that the mean values of the distributions are not representative of the distributions themselves. Or, in a practical way, observational data cannot be compared with the mean values provided by synthesis model to make physical inferences. This situation is especially relevant in the computation of total stellar masses of under-sampled clusters like the ones inferred to relate the total mass of a cluster with the mass of its more massive star. In these cases, one has to use all observational constraints (such as the number of stars with different masses), run a large enough set of Monte Carlo simulations which fulfills observational constrains, and obtain the distribution of possible total masses. Any other methodology (such as to complete the unobserved mass range up to a given mass with a mean value of a truncated IMF) introduce biased results (just like an artificial sorted sampling).

Finally, we show that the basic assumption implicit in the formalism of the IGIMF, that is, that when the integral of the IMF over some  given interval ($m^*({\cal{M}}_\mathrm{clus})$ to $m_\mathrm{up}$) is
normalised to unity  this provides the cluster mass ${\cal{M}}_\mathrm{clus}$ that contains just one star in that interval (and that this individual star has a mass equal to  $m^*({\cal{M}}_\mathrm{clus})$) is not correct. Extreme value theory shows that such an assumption just provides a ${\cal{M}}_\mathrm{clus}$ value which has a 63\% of probability to have a star with a mass {\it larger} than  $m^*({\cal{M}}_\mathrm{clus})$. In this situation some of the basis of the formulation of the IGIMF theory must be revised.

\acknowledgements MC thanks Sandro Bressan for showing him explicitly  
the non-IMF dependence of the high tail of the stellar luminosity function. He also thanks Alberto Buzzoni, Valentina Luridiana and Roberto and Elena Terlevich for several aspects related with sampling effects and the Initial Cluster Mass Function, and Fernando (and Mireia) Selman for fruitful discussions during the workshop on different aspects of science and life.  He also thanks who discovered him Albert Tarantola works in the inverse problem for fruitful conversations before, during and after the workshop. MC acknowledges the LOC for the help offered at all stages of the workshop. This work was supported by the Spanish {\it Programa Nacional de Astronom{\'{\i}}a y Astrof{\'{\i}}sica} through the project AYA2007-64712
and by the French ANR (09-BLAN-0228, POMMME).


\begin{thebibliography}{}
\expandafter\ifx\csname natexlab\endcsname\relax\def\natexlab#1{#1}\fi
\expandafter\ifx\csname url\endcsname\relax
  \def\url#1{\texttt{#1}}\fi
\expandafter\ifx\csname urlprefix\endcsname\relax\def\urlprefix{URL }\fi
\providecommand{\eprint}[2][]{\url{#2}}

\bibitem[{Barker et~al.(2008)Barker, de~Grijs, \& Cervi{\~n}o}]{BdGC08}
Barker, S., de~Grijs, R., \& Cervi{\~n}o, M. 2008, A{\&}A, 484, 711

\bibitem[{Buzzoni(1989)}]{Buzz89}
Buzzoni, A. 1989, Ap.J.S.S., 71, 817

\bibitem[{Cervi{\~n}o \& Luridiana(2004)}]{CL04}
Cervi{\~n}o, M., \& Luridiana, V. 2004, A{\&}A, 413, 145

\bibitem[{Cervi{\~n}o \& Luridiana(2005)}]{CL05}
--- 2005, in Resolved Stellar Populations, edited by .~M.~C. D.~Valls-Gabaud
  (ASP Conf. Ser.), in press. \eprint{arXiv:astro-ph/0510411}

\bibitem[{Cervi{\~n}o \& Luridiana(2006)}]{CLCL06}
--- 2006, A{\&}A, 451, 475

\bibitem[{Gumbel(1958)}]{Gum58}
Gumbel, E. J.~. 1958, Statistics of Extremes (New York: Columbia University
  Press)

\bibitem[{{Hernandez} \& {Valls-Gabaud}(2008)}]{Hernandez:2008qy}
{Hernandez}, X., \& {Valls-Gabaud}, D. 2008, \mnras, 383, 1603

\bibitem[{Marigo \& Girardi(2001)}]{MG01}
Marigo, P., \& Girardi, L. 2001, A{\&}A, 377, 132

\bibitem[{Massey(2010)}]{Mass10}
Massey, P. 2010, in UP: Have Observations Revealed a Variable Upper End of the
  Initial Mass Function? (ASP Conf. Ser.), in press.
  \eprint{arXiv:astro-ph/1008.1014}

\bibitem[{Oey \& Clarke(2005)}]{OC05}
Oey, M., \& Clarke, C. 2005, ApJ, 620, L43

\bibitem[{Pflamm-Altenburg \& Kroupa(2008)}]{P-AK08}
Pflamm-Altenburg, J., \& Kroupa, P. 2008, Nature, 455, 641

\bibitem[{Salpeter(1955)}]{Sal55}
Salpeter, E.~E. 1955, ApJ, 121, 161

\bibitem[{Selman \& Melnick(2008)}]{SM08}
Selman, F., \& Melnick, J. 2008, ApJ, 689, 816

\bibitem[{Sornette(2004)}]{Sor04}
Sornette, D. 2004, Critical phenomena in Natural Sciences (Berlin: Springer
  series in Synergetics)

\bibitem[{Tarantola(2006)}]{Tarantola}
Tarantola, A. 2006, Nature Physics, 2, 492

\bibitem[{Tinsley(1980)}]{Tin80}
Tinsley, B. 1980, Fundamentals of Cosmic Physics, 5, 287

\bibitem[{{Valls-Gabaud}(2011)}]{dvg2011}
{Valls-Gabaud}, D. 2011, in Local Group Cosmology, Proceedings of the XX Canary
  islands Winter School, edited by D.~{Martinez-Delgado} (Cambridge: Cambridge
  University Press)

\bibitem[{van Albada(1968)}]{vAlb69}
van Albada, T.~S. 1968, Bulletin of the Astronomical Institutes of the
  Netherlands, 20, 57

\bibitem[{Weidner et~al.(2010)Weidner, Kroupa, \& Bonnell}]{WKB10}
Weidner, C., Kroupa, P., \& Bonnell, I. A.~D. 2010, M.N.R.A.S., 401, 275

\end{thebibliography}

\end{document}